\documentclass[preprint,12pt]{elsarticle}

\usepackage{amssymb}
\usepackage{amsmath,amsfonts,algorithmic}
\usepackage{multirow}

\journal{Sensors and Actuators A: Physical}

\begin{document}

\begin{frontmatter}

\title{Classifying herbal medicine origins by temporal and spectral data mining of electronic nose}

\author[inst1]{Li Liu}
\author[inst1]{Xianghao Zhan}
\author[inst1]{Ziheng Duan}
\author[inst3]{Yi Wu}
\author[inst1]{Rumeng WU}
\author[inst1]{Xiaoqing Guan}
\author[inst1]{Zhan Wang}
\author[inst1]{You Wang*}
\author[inst1]{Guang Li}

\affiliation[inst1]{organization={Institute of Cyber-syste and Control, Zhejiang University},
            city={Hangzhou},
            postcode={310007}, 
            country={China}}
            
\affiliation[inst3]{organization={Institute of Intellegent Agricultural Equipment,Zhejiang university},
            city={Hangzhou},
            postcode={310007},
            country={China}}

\begin{abstract}
The origins of herbal medicines are important for their treatment effect, which could be potentially distinguished by electronic nose system. Because the difference in the odor fingerprint of herbal medicines from different origins can be tiny, the discrimination of origins can be much harder than that of different categories.  Therefore, better feature extraction methods are significant for this task to be more accurately done, but there lacks systematic studies on different feature extraction methods and a standardized manner to extract features from e-nose signals upon which most researchers agree. To investigate the effectiveness of multiple feature engineering approaches, we classified different origins of three categories of herbal medicines with different feature extraction methods: manual feature extraction, mathematical transformation, deep learning. With 50 repetitive experiments with bootstrapping, we compared the effectiveness of the extractions with a two-layer neural network w/o dimensionality reduction methods (principal component analysis, linear discriminant analysis) as the three base classifiers. Compared with the conventional aggregated features, the Fast Fourier Transform (FFT) method and our novel approach (longitudinal-information-in-a-line) showed an significant accuracy improvement($p<0.05$) on all 3 base classifiers and all three herbal medicine categories, with the highest median classification accuracy 0.675 and 0.7 over 30 experiments. Two of the deep learning algorithm we applied also showed partially significant improvement: one-dimensional convolution neural network(1D-CNN) and a novel graph pooling based framework - multivariate time pooling (MTPool), with the highest median accuracy 0.75 and 0.65.
\end{abstract}

\begin{graphicalabstract}
\includegraphics{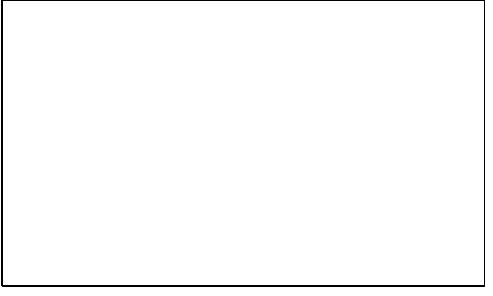}
\end{graphicalabstract}

\begin{highlights}
\item Different feature extraction methods were systematically investigated on classifying origins of herbal medicines.
\item The proposed long-line method includes both aggregated and temporal information, which outperformed the conventional aggregated method. 
\item Feature extraction method containing temporal information gained remarkable accuracy improvement with dimensionality reduction methods.
\end{highlights}


\end{frontmatter}


\section{Introduction}
Different alternative herbal medicines have distinct pharmaceutical values, because of not only different categories but also different geographical origins \cite{Herbal_medicine_origin}. The practictioners of herbal medicines attach significance to the specific medicines originated from the best-known origins \cite{herbal_medicine_introduction}. Therefore, the medicines from different geographical locations have a high variance in price, leaving space for frauds. Due to the tiny odor difference, there has been frauds in the herbal medicine markets that uses cheaper and inferior herbal medicine ingradients not originated from the most well-known origins \cite{chinese_herbal_medicine,herbal_medicine_origins}. To get the better treatment effect for patients, it is necessary to distinguish both the categories and geographical origins of the alternative herbal medicines. However, the similarities in appearances and odors make it difficult for experts to discriminate herbal medicines of the same category but from different origins. An accurate and cheap analytic method capturing the subtle differences is in need. \\
\indent Electronic nose (e-nose), has been proved to be effective and affordable in pattern recognition based on volatile organic compounds (VOCs). It has been successfully applied in lung cancer detection \cite{lungcancer}, dendrobiums identification \cite{Diao2020}, and herbal medicine category classification\cite{sensors12,IETBRAIN,augmentation}.\\
\indent Although the previous research showed prototypical success in classifying herbal medicine origins with a set of engineered features\cite{FeatureEngineering2019}, there is still much room for improvement. First, features extracted from e-nose signals significantly influences the classification performance. However, most previous publications in herbal medicine classification did not systematically compare the feature extraction methods and their respective effectiveness in terms of classification accuracy. As there is no universal features for e-nose agreed by researchers, several studies adopted the aggregated features and deemed them as a standard pipeline for e-nose pattern recognition \cite{FeatureEngineering2019,Wangzhan2017}. The aggregated features involved the steady-state and transient information\cite{vergara2012chemical} of the signals, which were mainly based on subjective domain-expert experience but only partially exploited the signal information without mining the temporal and spectral details encapsulated in the e-nose response signals. For example, Zhan et al. revealed the overabundance and low predictive power of some of the aggregated features \cite{FeatureEngineering2019}. Therefore, more universally applicable feature engineering methods worth further investigation. Second, in the previous publication to validate the feature extraction methods\cite{FeatureEngineering2019}, leave-one-out cross validation was used. The test data were used both in hyperparameter tuning and model evaluation. This might lead to overestimated accuracy. Furthermore, the previous study \cite{FeatureEngineering2019} did not verify the model robustness or test the statistical significance with repeated experiments.\\
\indent In this study, we improved our previous study on herbal medicine origin classification by systematically comparing different feature engineering methods with parallel experiments, to explore effective feature extraction methods with temporal and spectral information for higher classification accuracy. With a stricter model development and evaluation design, we also tested the statistical significance in the difference among various feature engineering approaches, which addressed the limited reproducibility issue in previous studies.\\

\section{Materials and Methods}

\subsection{Experiment and dataset}
\begin{figure*}[htbp]
\centerline{\includegraphics[width=\linewidth]{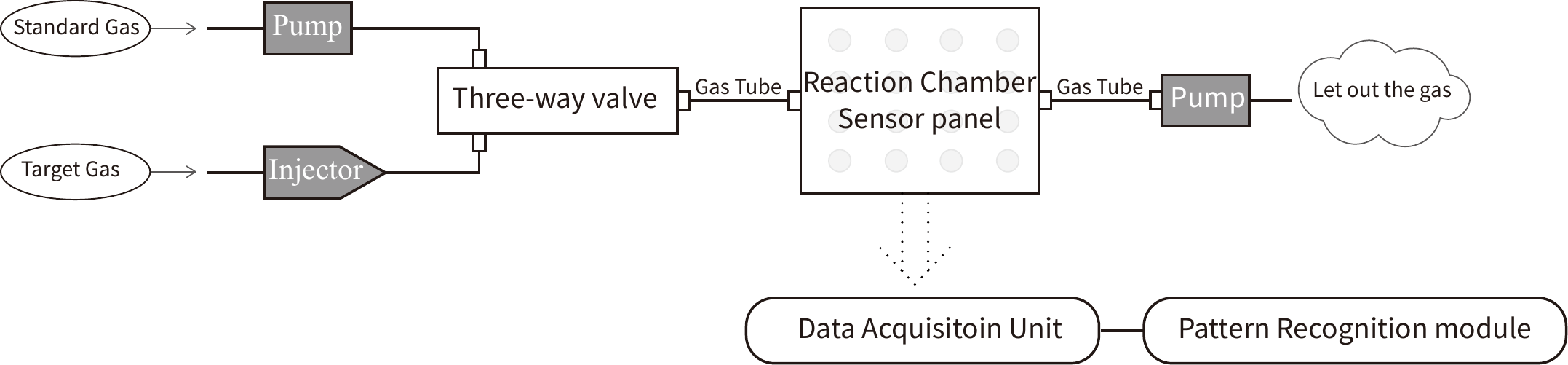}}
\label{e_nose_system_structure}
\caption{The brief description of the structure of electronic nose system. The e-nose system consists of four sub-systems: 1) the gas conveying system to control the flow, 2) the sensor reaction chamber where the target gas reacts the sensor panel, 3) the data acquisition unit for signals recording, and 4) the pattern recognition module for the data processing and classification.}
\end{figure*}

\begin{figure*}[htbp]
\centerline{\includegraphics[width=\linewidth]{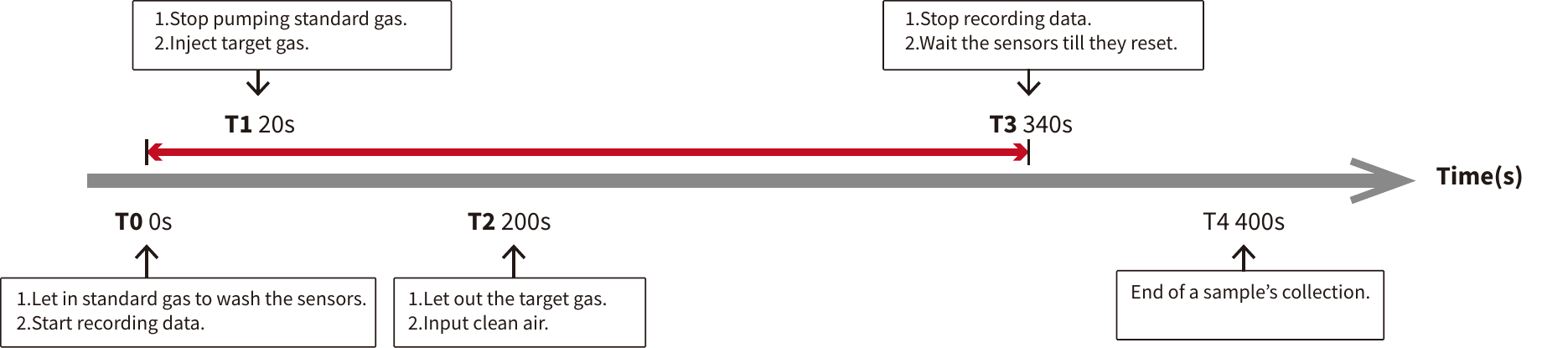}}
\label{data_collecting_timeline}
\caption{The description of the data collection process for a sample of target gas. For each one, the experiment process lasts for 400s, and the data recording interval (denoted by the red line) takes 340s. The data collection process includes 4 steps: 1) T0-T1: the initial standardization of sensors with clean air, when the sensor baseline values were recorded. 2) T1-T2: the reaction of target gas and the sensor panel. 3) T2-T3: the outflow of the target gas and the cleaning step of the reaction chamber. 4) T3-T4: the reset process of the system. }
\end{figure*}

In this study, we chose three categories of alternative herbal medicines as the study cases and collected three independent datasets: Radix Angelicae, Angelica Sinensis and Radix Puerariaem. Each dataset included 160 samples from 4 different origins \cite{FeatureEngineering2019}, as shown in Table \ref{Origin table}. The data was collected at the State Key Laboratory of Industrial Control Technology, Zhejiang University from Dec.2017 to Jan.2018 with a self-assembled e-nose system \cite{sensors12,FeatureEngineering2019,lungcancer,Wangzhan2017,Diao2020}. This self-assembled e-nose system is composed with four modules: the gas conveying system, the sensor reaction chamber where the gas reacts with sensors, the data acquisition unit, and the pattern recognition module for the data processing and classification. The structure of the e-nose system is shown in Fig. \ref{e_nose_system_structure}. The gas conveying system involves two gas pumps and a three-way valve to control the gas flow of the target gas and the standard gas (dry and clean air). 
The sensor reaction chamber consists of 16 metal-oxide semi-conductive (MOS) sensors: TGS (Taguchi Gas Sensors) from Figaro Engineering Inc, Osaka, Japan \cite{sensors12,IETBRAIN,FeatureEngineering2019}, and the specialities of 16 sensors are displayed in Table \ref{Sensors_speciality}. The data acquisition unit records the sensor reaction signals with a sampling rate of 100 Hz, with more details shown in Fig. \ref{data_collecting_timeline}. The entire process for a sample of target gas lasted for 400 seconds, and each sample of target gas stayed in the sensor reaction chamber for 180 seconds. The rest 220 seconds were for the sensors resetting process when we pumped the standard gas into the sensor reaction chamber for 20 seconds before pumping in the target gas and 200 seconds after pumping out the target gas. \\
\indent For each type of the medicines, the following steps were applied to extract the target gas :\\
\indent 1. Grinded the alternative herbal medicines into powders with an electrical pulverizer.\\
\indent 2. Took 8 grams of the power and put it into a 125 ml glass jar, and then used a para-film to seal the jar.\\
\indent 3. Heated the sample powders for 10 hours in a 50 ${ }^{\circ} \mathrm{C}$ thermostatic chamber, and then enabled the volatile gases to diffuse in the glass jar for 10 more hours.\\
\indent 4. From the headspace of each glass jar, took 10 mL gas to be the target gas for each medicine sample.\\
\indent Our experiment was done at laboratory where the environment temperature of 22 - 27 ${ }^{\circ} \mathrm{C}$ and the humidity was kept between 50 \% - 70 \%.
\begin{table}[htbp]
\caption{Three categories of herbal medicine from different origins}
\label{Origin table}
\renewcommand\arraystretch{1.5}
\begin{center}
\scalebox{0.8}{
\begin{tabular}{ccccc}
\hline
\textbf{Categories} & \textbf{Origin 1} & \textbf{Origin 2} & \textbf{Origin 3} & \textbf{Origin 4} \\ \hline
Radix Angelicae                  & Anhui            & Sichuan          & Hubei            & Zhejiang         \\ 
Angelica Sinensis                & Shaanxi          & Gansu            & Hubei            & Sichuan          \\ 
Radix Puerariae                  & Sichuan          & Hubei            & Anhui            & Hunan            \\ \hline
\end{tabular}}
\end{center}
\vspace{-3mm}
\end{table}

\begin{table}[]
\caption{Different sensors's particular high-sensitivity for volatile organic compounds.}
\label{Sensors_speciality}
\scalebox{0.8}{
\begin{tabular}{ccl}
\hline
No. & The Type Of Sensors & Individual Response Sensitivity                                                                                       \\ \hline
1   & TGS800      & Carbon monoxide, ethanol, methane, hydrogen, ammonia                                                                \\
2   & TGS813      & Carbon monoxide, ethanol, methane, hydrogen, isobutane                                                              \\
3   & TGS813      & Carbon monoxide, ethanol, methane, hydrogen, isobutane                                                              \\
4   & TGS816      & Carbon monoxide, ethanol, methane, hydrogen, isobutane                                                              \\
5   & TGS821      & Carbon monoxide, ethanol, methane, hydrogen                                                                         \\
6   & TGS822      & \begin{tabular}[c]{@{}l@{}}Carbon monoxide, ethanol, methane, acetone, n-hexane,\\ benzene, isobutane\end{tabular}  \\
7   & TGS822      & \begin{tabular}[c]{@{}l@{}}Carbon monoxide, ethanol, methane, acetone, n-Hexane,\\  benzene, isobutane\end{tabular} \\
8   & TGS826      & Ammonia, trimethyl amine                                                                                            \\
9   & TGS830      & Ethanol, R-12, R-11, R-22, R-113                                                                                    \\
10  & TGS832      & R-134a, R-12 and R-22, ethanol                                                                                      \\
11  & TGS880      & Carbon monoxide, ethanol, methane, hydrogen, isobutane                                                              \\
12  & TGS2620     & Methane, Carbon monoxide, isobutane, hydrogen                                                                       \\
13  & TGS2600     & Carbon monoxide, hydrogen                                                                                           \\
14  & TGS2602     & Hydrogen, ammonia ethanol, hydrogen sulfide, toluene                                                                \\
15  & TGS2610     & Ethanol, hydrogen, methane, isobutane/propane                                                                       \\
16  & TGS2611     & Ethanol, hydrogen, isobutane, methane                                                                               \\ \hline
\end{tabular}}
\end{table}

\indent
\subsection{Feature extraction}
\subsubsection{Manual extraction methods}
\begin{figure}[htbp]
\centerline{\includegraphics[width=0.8\linewidth]{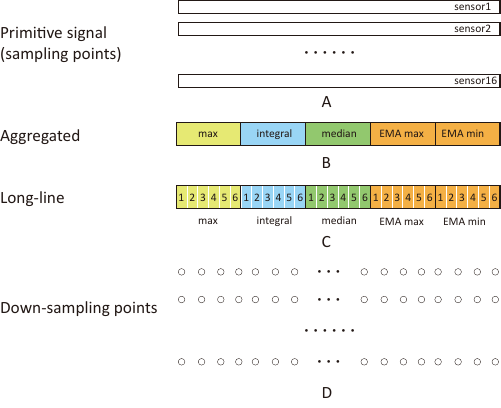}}
\caption{The illustration of the manual feature extraction processes.}
\label{feature extraction}
\vspace{-5mm}
\end{figure}

The manual feature extraction methods are shown in Fig. \ref{feature extraction} and Table. \ref{Feature type table}. The primitive signal contained the 16-sensor responses in 318 seconds with a sampling rate of 100Hz, with baseline already removed from these signals:
\begin{equation}
V=V_{S}-V_{0}
\end{equation}
In Fig. \ref{feature extraction} A, each bar denotes a temporal response of one sensor. We firstly introduced the sampling rate as a variable in feature engineering: as the signals were not changing rapidly, we used the down-sampling method with a sampling rate of 1Hz to reduce dimensionality. Then, we took the aggregated feature extraction method which extracted 5 features from each sensor, including:\\
\indent 1. The maximum voltage:
\begin{equation}
V_{\max }=\max (|V|)
\end{equation}
\indent 2. The integral value of voltage: 
\begin{equation}
V_{i n t}=\int_{0}^{T} V(t) d t
\end{equation}

\indent 3. The median of the temporal data series:
\indent \begin{equation}
V_{median }=median (V)
\end{equation}
\indent 4-5. The maximum and minimum value of the exponential moving average (EMA) of the derivative of voltage, with $\alpha = 1/SR$, $SR=100$:
\begin{equation}
E_{a}(V)=[\min (y(k)), \max (y(k))], 2000<k<34,000
\end{equation}. 
\begin{equation}
y(k)=(1-a) y(k-1)+a(V(k)-V(k-1))
\end{equation}
\begin{equation}
a= \frac{1}{S R}
\end{equation}
the effective of these aggregated features were proved effective in the previous publication\cite{FeatureEngineering2019,sensors12}. Then, we introduced a new method derived from the aggregated feature extraction method: the longitudinal-information-in-a-line (long-line) method. It was a variant version of the aggregated method proposed by us, with the aggregated features extracted from 6 separate time windows instead of the entire time range. In Fig. \ref{feature extraction} B, each cell (e.g. max) denotes an array of the results extracted from 16 sensors. In fig.\ref{feature extraction} C, each sub-cell (e.g. the sub-cell in max) denotes an array with results from 16 sensors.

\begin{table*}[htbp]
\caption{The feature extraction methods and classification algorithms used in this study.}
\renewcommand\arraystretch{1.2}
\label{Feature type table}
\begin{center}
\scalebox{0.6}{
\begin{tabular}{cccc}
\hline
\textbf{Feature extraction types}              & \textbf{Method}             & \textbf{Dimensionality reduction method} & \textbf{Classifier} \\ \hline
\multirow{2}{*}{Manual Extraction}             & Aggregated                  & PCA, LDA                                 & DNN                 \\
                                               & long-line                    & PCA, LDA                                 & DNN                 \\ \hline
\multirow{2}{*}{Signal Sampling}               & Sampling points                    & PCA, LDA                                 & DNN                 \\
                                               & down-sampling points               & PCA, LDA                                 & DNN                 \\ \hline
\multirow{2}{*}{Time-frequency Transformation} & Fast Fourier transformation & PCA, LDA                                 & DNN                 \\
                                               & Scalogram                   & PCA, LDA                                 & VGG+DNN                 \\ \hline
\multirow{3}{*}{Deep Learning}                 & \multicolumn{3}{c}{LSTM}                                                                     \\
                                               & \multicolumn{3}{c}{1D-CNN}                                                                   \\
                                               & \multicolumn{3}{c}{MTPool}                                                                      \\ \hline
\end{tabular}
}
\vspace{-5mm}
\end{center}
\end{table*}

\subsubsection{Mathematical transformation}
To extract spectral features, we applied two time-frequency transformations (shown in Table. \ref{Feature type table}): Fast Fourier transform (FFT) and scalogram analysis. The FFT method extracts the spectral densities of a signal, which decomposes the original sequence into several components with different frequencies\cite{FFT,FFT_2}. Scalogram uses continuous wavelet transform filter bank (implemented with MATLAB R2020a) to decompose signals and display the spatio-temporal information of a signal with the graph format\cite{cwt}. In this study, the sampling signals in Fig.\ref{feature extraction} A was used as the transformation input. We took the magnitudes of FFT results. Because after inspecting the spectral density distribution, we found the main frequency components gathered in 0-0.5HZ. We then used 10 frequency windows with a length of 0.05Hz to extract the spatial density information. The maximum, mean and median values in each window were collected as the featrues. The scalogram method requires image recognition method to extract numeral features from the scalograms, and therefore, we adopted pre-trained large convolutional neural network - VGG16\cite{scalogram_vgg,VGG_intro} to preprocess the scalogram features and condense the spatial-temporal information involved in the scalograms. To extract the coarse and detailed information of scalograms such as the edges and lines, we chose VGG's fully connected layer 7 as the output before further processing.

\subsubsection{Deep learning algorithms}
To explore the data-driven feature extraction methods, we employed three deep learning algorithms (shown in Table.\ref{Feature type table}): recurrent neural network with long short-term memory (LSTM), one-dimensional convolution neural network (1D-CNN), a newly proposed graph pooling based framework (MTPool) specific for multi-sensor data pattern recognition. Those algorithms can learn parameters to extract features from the primitive signals in a data-driven manner. Considering the size of our dataset, the computaional cost and to avoid overfitting, we chose the down-sampling points in Fig.\ref{feature extraction} D as the input. LSTM is a recurrent neural network (RNN) suited for time series data, and it is featured with the connection gates to utilize the information in previous state (also called memories)\cite{LSTM,LSTM2}. We designed an LSTM layer followed by a linear layer for further information processing, which was further connected to the output layer. As for the 1D-CNN approach, it is a special deep neural network which uses a 1-D convolution kernel to generate the information in a graph, and the kernel only moves in time rather than across sensors \cite{DNN,cnn_introduction} to utilize the temporal relationship of each sensor. After the convolution operation, three fully connected layers were employed to further process the information and give the output. 
MTPool is a novel graph pooling based framework\cite{Duan_MTPool}, which uses pairwise dependencies of multivariate time series to refine the nodes \cite{xu2020multivariate},\cite{wang2020mthetgnn}.
The proposal of MTPool is based on the fact that current deep learning methods for MTSC have two limitations:
(1) models which depend on the convolutional or recurrent neural networks cannot explicitly model the pairwise dependencies among variables;
(2) current spatial-temporal models based on GNNs are inherently flat and cannot hierarchically aggregate hub data \cite{ying2018hierarchical}. 
To overcome these challenges,  MTPool views the MTSC task as a graph classification task and use a graph pooling-based method to deal with it.
MTPool first converts MTS slices to graphs to attain the spatial-temporal features with graph structure learning and temporal convolution. 
In addition, MTPool utilizes a novel graph pooling strategy, where an 'encoder-decoder' mechanism is used to determine adaptive clustering centroids for cluster assignments.
Then GNNs and graph pooling layers are used for joint graph representation learning and graph coarsening, after which the graph is progressively coarsened to one node. 
At last, a differentiable classifier takes this coarsened one-node graph as input to get the final predicted class.
What must be mentioned here, the hyperparameters of the deep learning network structure, such as the number of fully connected layers after the LSTM layer and the 1-D convolution layer, were tuned based on the validation data, which will be introduced in the next subsection. 


\subsection{Prediction and evaluation protocol}\label{AA}
To avoid overestimating the classification accuracy, we adopted the hold-out test protocol instead of the 'leave-one-out' validation protocol which was used in our previous study \cite{FeatureEngineering2019}: the test set for model evaluation was independent of the training and hyperparameter tuning processes. For each category in Table \ref{Origin table}, we randomly selected 120 samples to be the training set, leaving the rest 40 for testing.\\
\indent For manual extraction and mathematical transformation methods, because the feature extraction was independent of the class labels, we firstly applied the feature extraction methods on the entire dataset, and then partitioned the entire dataset into the training set and the test set. In this study, we chose DNN as the basic classifier because it is flexible and does not require strict model and data distribution assumptions. 

Because dimensionality reduction is commonly used in many classification studies, as we are comparing the effectiveness of different feature extraction, we also evaluated the classification performances of these feature extraction methods in combination with the dimensionality reduction approaches for a more comprehensive conclusion. We employed 2 dimensionality reduction methods: principal component analysis (PCA), linear discriminant analysis (LDA), and combine them with the base classifier DNN to form the following three base classifiers: DNN, PCA-DNN, LDA-DNN. To tune the hyperparameters of the classifiers, we used a 5-fold cross validation protocol within the training set. The hyperparameters included: the length of width of the fixed spectral window in the FFT method, the reduced dimensionality in PCA and LDA, the image processing network (VGG or Inception) and its output layer, the number of hidden layers and number of hidden units of DNN, and the number of fully connected layers and the hidden units of 1D-CNN. To test the statistical significance in the difference in accuracy of various feature engineering methods, we repeated the experiment for 50 times by bootstrapping the 120 training samples with different random seeds, and then we recorded the prediction accuracy on the test set. With the fifty results, we performed Wilcoxon signed-rank tests to test statistical significance. The paired t-test was not used in this study, because the Shapiro-Wilk tests rejected the normality assumption of some groups of accuracy results and Wilcoxon signed-rank does not rely on the normal distribution assumption.\\
\indent For the deep learning methods, as the training and validation took a longer time, we directly partitioned the entire dataset into 80/40/40 for train/validation/test sets. The validation set was used for hyperparameter tuning process, which included the convolution kernel size and the network structure in 1D-CNN, the hidden unit dimensionality, number of units in the linear layer after the recurrent layer, learning rate and training epochs in LSTM, the number of centroid heads and pooling layers in MTPool. With the hyperparameters tuned, we retrained the deep learning models using training set in the same bootstrapping pipeline, and tested its prediction accuracy performance on test set. We took the feature extraction method used in previous studies (aggregated features) as the baseline \cite{FeatureEngineering2019,sensors12}, and applied Wilcoxon signed-rank tests to test the statistical significance.

\section{Results}
\indent The accuracies of different feature extraction methods in the herbal medicine origin classification are shown in Fig. \ref{result}. For Radix Angelicae, except for the scalogram with VGG network, all the other feature extraction methods outperformed baseline on three classifiers($p<0.05$). The highest median classification accuracy was 0.775 reached by LSTM. The signal sampling methods(down-sampling points and sampling points) generally manifest a higher accuracy improvement than manual extraction methods (baseline and long-line), when working on a same classifier.\\
\indent For Angelica Sinensis, the long-line, down-sampling, sampling and FFT methods outperformed the baseline when working on a classifier($p<0.05$). For deep learning methods, only 1D-CNN outperformed the baseline (median accuracy: 0.625). \\
\indent For Radix Puerariae, long-line and FFT methods performed better than baseline on all classifiers. The best performance was from sampling points with LDA-DNN as the classifier (median accuracy: 0.725).\\

\begin{figure*}[htbp]
\centerline{\includegraphics[width=0.8\linewidth]{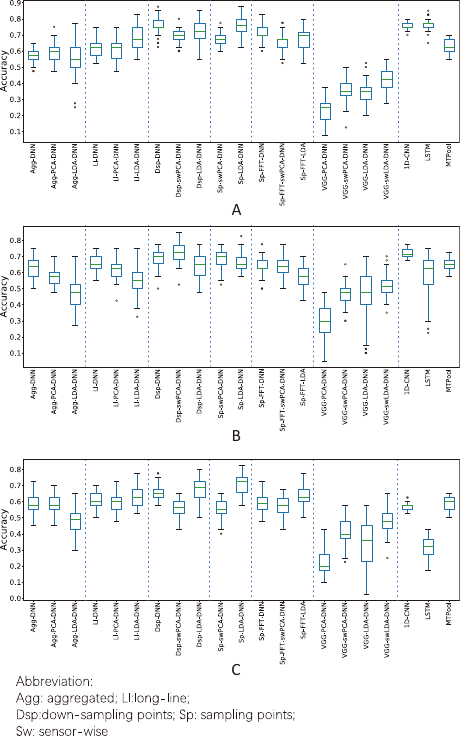}}
\label{result}
\caption{The classification accuracy of different feature extraction methods on classifying herbal medicine origins for three categories(A: Radix Angelicae, B: Angelica Sinensis, C: Radix Puerariae)}
\end{figure*}


\begin{figure*}[htbp]
\centerline{\includegraphics[width=\linewidth]{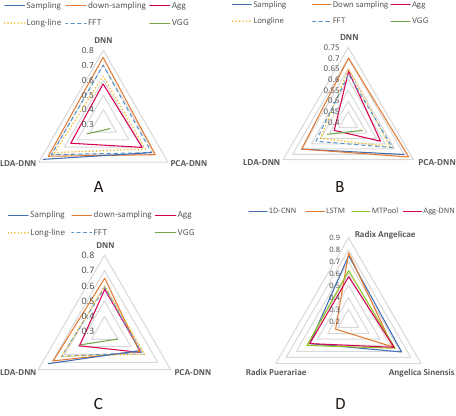}}
\label{result}
\caption{The comparison of the corresponding median values of the classification accuracy of different feature extraction methods on classifying herbal medicine origins for three categories(A: Radix Angelicae, B: Angelica Sinensis, C: Radix Puerariae). Specially, the results of deep learning algorithms are shown in D.}
\end{figure*}

\section{Discussion}
\indent Compared with the conventional aggregated features, the long-line and FFT feature extraction methods generally improved the classification accuracy with three classifiers on classifying different origins for all three different categories of herbal medicines. Furthermore, two feature extraction methods always performed better than baseline when we set the classifier as a fixed variable: the down-sampling points with DNN and LDA-DNN, sampling points methods with LDA-DNN. As for the deep learning methods, 1D-CNN and MTPool showed a general improvement on the origin classification tasks for three different categories of herbal medicines, which also manifested the power of deep learning. Compared with deep learning, the manual feature extraction methods (such as long-line, FFT) still show better effectiveness on our dataset. This is probably due to the small number of our samples, which may lead to the overfitting in the complicated deep learning models.
\\
\indent Based on the results and comparison among different feature engineering approaches, we provide the users who are willing to classify herbal medicine origins with e-nose with the following general suggestions on feature extraction strategies: with undetermined classifiers, for higher robustness in high classification accuracy, the long-line method, and the FFT approach are preferred than the conventional aggregated method; To pursue a faster feature extraction and classification process, down-sampling points and sampling points with LDA-DNN are recommended. \\
\indent This study showed more convenient and efficient feature extraction methods than the aggregated features which have been used in many previous studies. However, in the experiments, the advantages of those methods might slightly vary when they are adopted on different categories of herbal medicines. This was probably due to the limitation of our small-scale dataset, which may lead to variance in the results. Therefore, further validation will be done on larger datasets with more categories of herbal medicines originated from different regions.

Besides the comprehensive analysis and comparison of feature extraction methods in this dataset, we published this dataset with the feature engineering methods we used (aggregated features, long-line, down-sampling points, sampling points, FFT) on Github (https://github.com/xzhan96-stf/Herbal-medicine-origin-e-nose) for researchers to develop better algorithms, which can be compared with the results in this study as the benchmark.

\section{Conclusion}
In this study, different feature extraction methods were investigated in the classification tasks where herbal medicine origins were discriminated, with the primitive signals collected by an e-nose system. To get a more credible conclusion, those approaches were individually tested on three categories of herbal medicine (each one has 4 origins) as three independent dataset. The feature extraction methods fall into four categories: signal sampling, manual extraction methods, mathematical transformation, and deep learning algorithms. Working with three basic classifiers ( DNN, LDA-DNN, PCA-DNN), the effectiveness of these approaches were compared. In general, with a fixed clssifier, the extraction methods with both temporal information and expert knowledge (such as longline and FFT) performs well and robustly. However, when working with a strong dimensionality reduction method (such as LDA, down sampling), the extraction methods containing more temporal information (such as down-sampling points and sampling points with LDA-DNN) can sometimes reach remarkably high accuracy. As there is no standard feature extraction pipeline for mining electronic nose data, this study provides the research community with a general guidance on the potential better choice of features when work with electronic nose to improve the accuracy of pattern recognition.

\bibliographystyle{IEEEtran} 
\bibliography{cited}

\end{document}